# Spin relaxation in Kondo lattices

S I Belov, A S Kutuzov and B I Kochelaev

Kazan Federal University, Kremlevskaya, 18, Kazan 420008, Russian Federation

E-mail: Sergei.Belov@ksu.ru

**Abstract.** A model of spin relaxation in Kondo lattices is proposed to explain the presence of an electron spin resonance (ESR) signal in the heavy fermion compounds $YbRh_2Si_2$ and $YbIr_2Si_2$. Coupled equations for dynamical susceptibilities of Kondo ions and conduction electrons are derived by means of the functional derivative method. The perturbational scaling approach reveals the collective spin motion of Yb-ions with conduction electrons in the bottleneck regime. A common energy scale due to the Kondo effect regulates the temperature dependence of the different kinetic coefficients and results in a mutual cancelation of all divergent parts in a collective spin mode. The angular dependence of the ESR linewidth is shown to be in a qualitative agreement with experimental data on $YbRh_2Si_2$ and $YbIr_2Si_2$. Linewidth contributions other than the Kondo interaction are also discussed.

## 1. Introduction

The discovery of electron spin resonance (ESR) in the Kondo lattice $YbRh_2Si_2$ below the Kondo temperature $T_K = 25$ K [1] ($T_K$ revealed by specific heat data) was a great surprise for the condensed matter physics community and stimulated different approaches for its explanation [2, 3, 4]. According to the picture based on the single ion Kondo effect, magnetic moments of Yb-ions are screened by conduction electrons at $T < T_K$ and the ESR linewidth is of the order $\Delta \nu = k_B T_K / 2\pi\hbar = 500$ GHz. The experimental results were completely opposite: at X-band (9.4 GHz) and $T = 0.7$ K a linewidth of 0.3 GHz was observed. The angular dependence of the ESR $g$-factor and linewidth reflects the local properties of Yb-ion in a crystal electric field (CEF). Similar results were obtained later for $YbIr_2Si_2$ [5]. It seems that many properties of heavy fermion systems in the non-Fermi-liquid state can be described in terms of quasi-localized $f$-electrons. Recently we successfully studied the static magnetic susceptibility of $YbRh_2Si_2$ and $YbIr_2Si_2$ at temperatures below $T_K$ [6], based on entirely local properties of Yb-ions in a CEF. In another work [4] a theoretical model of spin relaxation was proposed to explain the ESR signal existence in Kondo lattice systems with heavy fermions. We showed that the collective spin motion of quasi-localized $f$-electrons and wide-band conduction electrons is the key ingredient for understanding this phenomenon. Our model successfully explained the temperature and magnetic field dependence of the ESR linewidth and resonance frequency in the case of the static magnetic field oriented perpendicular to the crystal symmetry axis.

This work gives a more detailed account of our approach developed in [4] and extends the earlier results to the case of an arbitrary angle between the static magnetic field and the crystal symmetry axis. In section 2 we analyze the anisotropy of the Kondo interaction in the heavy fermion systems $YbRh_2Si_2$ and $YbIr_2Si_2$ and obtain the basic Hamiltonian of our model. Section 3 presents the model of spin relaxation in a Kondo lattice. Equations of spin motion are derived up to the third order in the Kondo exchange constants. In section 4 the perturbation expansion is improved by the renormalization of the Kondo couplings at low temperatures on the basis of the "Poor Man's Scaling" approach, and section 5 concludes the article. The mathematical basis of our approach is the functional derivative method of Kadanoff and Baym [7]. We outlined it briefly in the appendix.

## 2. Theoretical model

Our basic theoretical model includes the kinetic energy of conduction electrons, the Zeeman energy, the Kondo interaction of Yb-ions with conduction electrons, and the coupling between the Yb-ions via conduction electrons (RKKY interaction).

A free $Yb^{3+}$ ion has a $4f^{13}$ configuration with one term $^2F$. The spin-orbit interaction splits the $^2F$ term into two multiplets: $^2F_{7/2}$ with $J = 7/2$ and $^2F_{5/2}$ with $J = 5/2$, where $J$ is the value of the total momentum $\mathbf{J} = \mathbf{L} + \mathbf{S}$ with $\mathbf{L}$ and $\mathbf{S}$ as the orbital and spin momentum of the ion. The excited multiplet $^2F_{5/2}$ is separated from the ground state $^2F_{7/2}$ by about 1 eV. Since this value is much larger than the energy of the crystal electric field (CEF), we consider in the following the ground multiplet only. It is reasonable to express the Zeeman energy of the $i$-th Yb-ion $H_{ZJ}$ for the lowest multiplet $J = 7/2$ via the total electronic momentum of the ion. Using the Lande $g$-factor $g_J$, we have

$$H_{ZJ} = (\mathbf{L}_i + 2\mathbf{S}_i)\mathbf{B} = g_J \mathbf{J}_i \mathbf{B}, \qquad (1)$$

where $\mathbf{B}$ is the external magnetic field multiplied by the Bohr magneton.

The Kondo exchange coupling of the rare earth ion with the conduction electrons occurs due to the hybridization of their wave functions at the ion site. The exchange integral can be written in the form (see, e.g. [8])

$$A(\mathbf{k}, \mathbf{k}') = \sum_i \int \psi^*(\mathbf{r}', \mathbf{k}) \Psi_{4f}(\mathbf{r}_1, ..., \mathbf{r}_i, ..., \mathbf{r}_n) \frac{e^2}{|\mathbf{r}_i - \mathbf{r}'|} \psi(\mathbf{r}_i, \mathbf{k}') \Psi_{4f}(\mathbf{r}_1, ..., \mathbf{r}', ..., \mathbf{r}_n) d\mathbf{r}' d\mathbf{r}_1 ... d\mathbf{r}_n. \qquad (2)$$

Here $\psi(\mathbf{r}, \mathbf{k})$ is the Bloch wave function of the conduction electrons. The wave function of the 4$f$-electrons $\Psi_{4f}$ is represented by the determinant constructed from the one-electron wave functions of the type $R_{4f}(r_i) Y_3^m(\mathbf{r}_i)$. Expanding the Bloch functions and $|\mathbf{r}_i - \mathbf{r}'|^{-1}$ in spherical harmonics, one can obtain $A(\mathbf{k}, \mathbf{k}')$ as an expansion in multipoles. As a matter of fact the small parameter of this expansion is the value $k_F \langle r_{4f} \rangle \ll 1$, the product of the wave vector of the conduction electron at the Fermi surface and the average radius of the 4$f$-electron. The Kondo interaction at the $i$-th site corresponding to the zero order of this expansion is isotropic and can be written in the form [9, 10]

$$H_{KJ} = A_0 \mathbf{S}_i \boldsymbol{\sigma}_i = A_0 (g_J - 1) \mathbf{J}_i \boldsymbol{\sigma}_i, \qquad (3)$$

where $\boldsymbol{\sigma}_i$ is the spin density of the conduction electrons at the ion site. The next terms of the expansion in multipoles are less and anisotropic (detailed calculation of them can be found in [11, 12] and especially in [13]). In the following we neglect these terms, since the CEF gives a very strong anisotropy for the zero order of the Kondo exchange interaction (3).

The tetragonal CEF splits the ground multiplet into four Kramers doublets, each one described by the wave functions of the type $\psi_\pm = \sum_M c_{\pm M} |\pm M\rangle$. Within the every Kramers doublet the total Zeeman energy can be represented by the effective spin Hamiltonian with an effective spin $S = 1/2$:

$$H_{ZS} = \sum_i \left[ g_\perp \left( S_i^x B_i^x + S_i^y B_i^y \right) + g_\parallel S_i^z B_i^z \right], \qquad (4)$$

where $g_\perp = g_J \langle \psi_+ | J_x + i J_y | \psi_- \rangle$ and $g_\parallel = 2 g_J \langle \psi_+ | J_z | \psi_+ \rangle$ (details see in [6, 14, 15]).

According to neutron scattering experiments [16, 17] the excited energy levels are located at 17, 25, 43 meV (197, 290, 499 K) in the case $YbRh_2Si_2$ and at 18, 25, 36 meV (209, 290, 418 K) for $YbIr_2Si_2$. It means that the physics of low energy spin excitations at temperatures $T \ll 200\,\mathrm{K}$ can be described by the ground Kramers doublet. In the following we shall relate the Zeeman Hamiltonian (4) to the ground Kramers doublet; in other words, we have projected to this state the starting Zeeman energy with the isotropic Lande $g$-factor.

It is evident that the projection of the isotropic Kondo interaction (3) involves the same matrix elements of the total momentum **J** as in the case of the Zeeman energy. Hence, after projection onto the ground Kramers state the total Kondo interaction can be expressed via the $g$-factors $g_\perp$, $g_\parallel$ given above:

$$H_{s\sigma} = \sum_i \left[ J_\perp \left( S_i^x \sigma_i^x + S_i^y \sigma_i^y \right) + J_\parallel S_i^z \sigma_i^z \right] \qquad (5)$$

with $J_\parallel = Jg_\parallel$, $J_\perp = Jg_\perp$, $J = A_0 \dfrac{g_J - 1}{g_J}$. As a matter of fact all these results are simple consequences of the well known the Wigner-Eckart theorem.

The same arguments can be used to reveal the anisotropy of the RKKY interaction between the Kondo ions. Although this interaction appears in the second order of the Kondo interaction (3) it is convenient to consider it independently. The result is given, in particular, in [8]. Starting with the isotropic exchange Hamiltonian for two Kondo ions $H_{ex}^{ij} = I_{ij}^{RKKY} \mathbf{S}_i \mathbf{S}_j = I_{ij}^{RKKY}(g_J - 1)^2 \mathbf{J}_i \mathbf{J}_j$ the authors of [8] arrive to the effective anisotropic interaction

$$H_{RKKY} = \sum_{ij} I_{ij} \left[ g_\perp^2 \left( S_i^x S_j^x + S_i^y S_j^y \right) + g_\parallel^2 S_i^z S_j^z \right], \qquad (6)$$

where $I_{ij} = I_{ij}^{RKKY} \left( \dfrac{g_J - 1}{g_J} \right)^2$. It is evident that the same procedure was used as in deriving the equation (5).

The kinetic energy of the conduction electrons and their Zeeman energy can be written as

$$H_c = \sum_{ij\lambda} t_{ij} c_{i\lambda}^+ c_{j\lambda} - \mu \sum_{i\lambda} c_{i\lambda}^+ c_{i\lambda}, \qquad (7)$$

$$H_{z\sigma} = g_\sigma \sum_i \mathbf{B} \boldsymbol{\sigma}_i. \qquad (8)$$

Here $\lambda = \pm 1$ labels the orientation of the conduction electron spin, $\mu$ is the chemical potential, and $g_\sigma$ denotes the $g$-factor of conduction electrons. The conduction electron density is expressed in terms of the creation and annihilation operators

$$\boldsymbol{\sigma}_i = \sum_{\lambda\lambda'} \mathbf{s}_{\lambda\lambda'} c_{i\lambda}^+ c_{i\lambda'}, \qquad (9)$$

where $\mathbf{s}_{\lambda\lambda'}$ are the matrix elements of the spin operators $s = 1/2$.

Collecting all terms together we obtain an effective Hamiltonian

$$H = H_c + H_Z + H_{S\sigma} + H_{RKKY}. \qquad (10)$$

Here $H_c$, $H_{s\sigma}$, $H_{RKKY}$ are defined by (7), (5), (6), respectively, $H_Z$ is the total Zeeman energy of conduction electrons and Yb-ions:

$$H_Z = H_{ZS} + H_{Z\sigma} = B \sum_i \left[ g_\sigma (\sin\theta\, \sigma_i^y + \cos\theta\, \sigma_i^z) + g_\perp \sin\theta\, S_i^y + g_\parallel \cos\theta\, S_i^z \right], \qquad (11)$$

where $\theta$ is the angle between the static magnetic field and the crystal symmetry axis.

The next step is to diagonalize, for convenience of further calculations, the Zeeman part of the effective Hamiltonian. After an unitary transformation to a new quantization axis the Hamiltonian (10) takes its final form $H = H_0 + H_{S\sigma} + H_{RKKY}$ with

$$H_0 = \sum_{ij\lambda} t_{ij} c_{i\lambda}^+ c_{j\lambda} + \sum_{i\lambda} \varepsilon^\lambda c_{i\lambda}^+ c_{i\lambda} + g_s B \sum_i S_i^z , \qquad (12)$$

$$H_{s\sigma} = \sum_{i,\alpha\beta} J_{\alpha\beta} \sigma_i^\alpha S_i^\beta , \qquad (13)$$

$$H_{RKKY} = \sum_{ij,\alpha\beta} I_{ij}^{\alpha\beta} S_i^\alpha S_j^\beta . \qquad (14)$$

Here $\varepsilon^\lambda = -\mu + \lambda g_\sigma B/2$, $\alpha,\beta = x,y,z$,

$$g_s = \sqrt{g_\parallel^2 \cos^2\theta + g_\perp^2 \sin^2\theta} , \qquad (15)$$

$J_{\alpha\beta}$, $I_{ij}^{\alpha\beta}$ represent Kondo- and RKKY couplings after rotation of the quantization axis:

$$\begin{aligned}
&J_{xx} = J g_\perp, \; J_{yy} = J g_\parallel g_\perp / g_s, \; J_{zz} = J g_s, \\
&J_{yz} = J \frac{g_\perp^2 - g_\parallel^2}{g_s} \sin\theta\cos\theta, \\
&J_{zy} = J_{xy} = J_{yx} = J_{xz} = J_{zx} = 0,
\end{aligned} \qquad (16)$$

$$\begin{aligned}
&I^{xx} = I g_\perp^2, \; I^{yy} = I \frac{g_\parallel^2 g_\perp^2}{g_s^2}, \; I^{zz} = I\left(g_\parallel^2 + g_\perp^2 - \frac{g_\parallel^2 g_\perp^2}{g_s^2}\right), \\
&I^{yz} = I^{zy} = I g_\parallel g_\perp \frac{g_\perp^2 - g_\parallel^2}{g_s} \sin\theta\cos\theta, \\
&I^{xy} = I^{yx} = I^{xz} = I^{zx} = 0.
\end{aligned} \qquad (17)$$

The angle $\psi$ between the new quantization axis and the crystal symmetry axis is the one given by the relation

$$\cot\psi = \frac{g_\parallel}{g_\perp} \cot\theta . \qquad (18)$$

### 3. Spin relaxation
The ESR response is given by the total transverse dynamical susceptibility

$$\chi(\omega) = \sum_{\alpha\beta} \chi_{\alpha\beta}(\omega); \quad \alpha,\beta = s,\sigma \qquad (19)$$

with partial susceptibilities $\chi_{\alpha\beta}(\omega)$:

$$\begin{aligned}
\chi_{ss} &= -g_\perp^2 \langle\langle S^- | S^+ \rangle\rangle, & \chi_{s\sigma} &= -g_\perp g_\sigma \langle\langle S^- | \sigma^+ \rangle\rangle, \\
\chi_{\sigma s} &= -g_\perp g_\sigma \langle\langle \sigma^- | S^+ \rangle\rangle, & \chi_{\sigma\sigma} &= -g_\sigma^2 \langle\langle \sigma^- | \sigma^+ \rangle\rangle.
\end{aligned} \qquad (20)$$

Here $\langle\langle A | B \rangle\rangle$ is the Fourier transform of a retarded Green function

$$\langle\langle A | B \rangle\rangle = -i \int_0^\infty dt \, \exp(i\omega t) \langle [A(t), B] \rangle , \qquad (21)$$

**S**, **σ** are the total spin operators of Yb-ions and conduction electrons, respectively, $S^\pm = (S^x \pm iS^y)/\sqrt{2}$, $\sigma^\pm = (\sigma^x \pm i\sigma^y)/\sqrt{2}$, $\langle...\rangle$ means the statistical average at temperature $T$. The definition (20), (21) imply the symmetry relation

$$\chi_{\alpha\beta}^*(\omega) = \chi_{\beta\alpha}(\omega^*) . \qquad (22)$$

$\chi_{\alpha\beta}$ can be represented as the analytical continuation of the temperature Green functions $A$, $B$, $D$ (see appendix, (A.8)) onto the real axis:

$$\chi_{ss}(\omega+i0) = -g_\perp^2 D^{-+}(\mathbf{k}=0, i\omega_n \to \omega+i0),$$
$$\chi_{\sigma\sigma}(\omega+i0) = -g_\sigma^2 A^{-+}(\mathbf{k}=0, i\omega_n \to \omega+i0),$$
$$\chi_{\sigma s}(\omega+i0) = -g_\sigma g_\perp B^{-+}(\mathbf{k}=0, i\omega_n \to \omega+i0),$$
$$\chi_{s\sigma}(\omega+i0) = \chi_{\sigma s}^*(\omega-i0).$$
(23)

The definitions, main properties and calculation technique for the temperature Green functions are given in appendix.

As a first step we find $\chi_{\alpha\beta}$ for the high temperature region $T > B$ up to the third order in the Kondo coupling. The RKKY interaction being treated in molecular field approximation only. The result is convenient to be presented as a matrix equation

$$\begin{pmatrix} a_{ss} & a_{s\sigma} \\ a_{\sigma s} & a_{\sigma\sigma} \end{pmatrix} \begin{pmatrix} \chi_{ss} & \chi_{s\sigma} \\ \chi_{\sigma s} & \chi_{\sigma\sigma} \end{pmatrix} = \begin{pmatrix} g_\perp^2 \langle S^z \rangle & 0 \\ 0 & g_\sigma^2 \langle \sigma^z \rangle \end{pmatrix}$$
(24)

with

$$a_{ss} = \omega - \omega_s + \Sigma_{ss}, \qquad a_{s\sigma} = g_\perp^2 \langle S^z \rangle \lambda - \frac{g_\perp}{g_\sigma} \Sigma_{s\sigma},$$
$$a_{\sigma s} = g_\sigma^2 \langle \sigma^z \rangle \lambda - \frac{g_\sigma}{g_\perp} \Sigma_{\sigma s}, \qquad a_{\sigma\sigma} = \omega - \omega_\sigma + \Sigma_{\sigma\sigma}.$$
(25)

Here $\lambda = J(g_\parallel + g_s)/2 g_s g_\sigma$, $\langle S^z \rangle = -g_s B/4(T+T_W) + O(J)$, $\langle \sigma^z \rangle = -g_\sigma \rho B/2 + O(J)$, $\omega_s$, $\omega_\sigma$ are resonance frequencies of Yb-ions and conduction electrons, respectively, containing the first order Knight shifts due to the Kondo- and RKKY interaction:

$$\omega_s = g_s B + J g_s \langle \sigma^z \rangle + 4 T_W \langle S^z \rangle,$$
$$\omega_\sigma = g_\sigma B + J g_s \langle S^z \rangle.$$
(26)

$T_W$ is the Weiss temperature which originates from the RKKY interaction in a molecular field approximation:

$$T_W = \frac{1}{4}\left( g_\parallel^2 + g_\perp^2 - \frac{g_\parallel^2 g_\perp^2}{g_s^2} \right) \sum_i I_{ij},$$
(27)

$\rho$ is the conduction electron density at the Fermi surface.

The coefficients $\Sigma_{\alpha\beta}$ describe the spin kinetics in the system of Yb-ions and conduction electrons. Their imaginary parts give the relaxation rates, their real parts give the corresponding resonance frequencies. We calculate $\Gamma_{\alpha\beta} \equiv \text{Im}(\Sigma_{\alpha\beta})$ up to the third order in the Kondo interaction at high temperatures $T > B$:

$$\Gamma_{ss}(\omega) = \frac{\pi}{2} \frac{\omega}{\omega_s} (\rho J)^2 T \left[ \frac{3}{2} g_\parallel^2 + g_\parallel^2 \left( 1 - \frac{g_\perp^2}{2 g_s^2} \right) - 4\rho J g_\parallel g_\perp^2 \ln(T/W) \right],$$

$$\Gamma_{\sigma\sigma}(\omega) = \frac{\pi}{4} \frac{\omega}{\omega_s} \frac{g_s}{g_\sigma} \rho J^2 \frac{T}{T+T_W} \left[ g_\perp^2 + \frac{g_\parallel^2 + g_s^2}{2} - 4\rho J g_\parallel g_\perp^2 \ln(T/W) \right],$$
(28)

$$\Gamma_{\sigma s}(\omega) = \frac{\pi}{2}\frac{\omega}{\omega_s}(\rho J)^2 g_\perp (g_\| + g_s)T\left[1 - \frac{g_\perp^2 + g_\| g_s}{g_s}\rho J \ln(T/W)\right],$$

$$\Gamma_{s\sigma}(\omega) = \Gamma_{\sigma s}(\omega)\frac{g_s}{2g_\sigma \rho (T + T_W)}.$$

Here $\omega$ is the frequency of the alternating transverse magnetic field, $W$ is the bandwidth of the conduction electrons and $g_s$ is defined by (15).

The kinetic coefficient $\Gamma_{ss}(\omega = \omega_s)$ represents the relaxation rate of the transverse magnetic moment of Yb-ions toward conduction electrons which remain in a thermodynamical equilibrium state (Korringa relaxation). For an isotropic system with $g_\| = g_\perp = g$ we have the well known result in the second order in $J$: $\Gamma_{ss} = \pi(\rho J g)^2 T$. The third order reveals logarithmic Kondo anomalies which can be important at low temperatures. Similarly, substituting $\omega = \omega_\sigma$ into $\Gamma_{\sigma\sigma}(\omega)$ we obtain the Overhauser relaxation rate (conduction electrons relax toward the Yb spin system, being in the equilibrium with the thermal bath). Note that the strong anisotropy of the Kondo interaction breaks the detailed balance relation for $\Gamma_{ss}(\omega_s)$, $\Gamma_{\sigma\sigma}(\omega_\sigma)$ except for the cases of parallel ($g_s = g_\|$) and perpendicular ($g_s = g_\perp$) orientation of the static magnetic field to the crystal symmetry axis.

However, an equilibrium state approximation for the conduction electron spin system is not valid to study the ESR response of the samples with a high concentration of Kondo ions [18] which is especially the case for a Kondo lattice. Instead, one has to treat the coupled kinetic equations (24) for both magnetic moments of Kondo ions and conduction electrons. Hence two additional kinetic coefficients $\Gamma_{s\sigma}$, $\Gamma_{\sigma s}$ couple the kinetic equations of motion of the transverse magnetizations of localized moments and conduction electrons. Besides, for a correct analysis of a stationary solution one has to take into account the spin relaxation of Kondo ions and conduction electrons spin toward the thermal bath ("lattice"). Correspondingly, the kinetic coefficients $\Sigma_{ss}$, $\Sigma_{\sigma\sigma}$ in equations (24) should be replaced with $\Sigma_{ss} + \Sigma_{sL}$, $\Sigma_{\sigma\sigma} + \Sigma_{\sigma L}$, respectively. To the second order in $J$ our results coincides with equations derived by means of the non-equilibrium statistical operator [19]. To study the ESR response of the total system we have to find solutions of the system (24), where the coefficients $a_{\alpha\beta}$ include the spin lattice relaxation terms. The poles of the total dynamical susceptibility are determined by the condition $a_{ss}a_{\sigma\sigma} - a_{s\sigma}a_{\sigma s} = 0$. The result are two complex roots: their real parts represent resonance frequencies, their imaginary parts represent the corresponding relaxation rates. We are interested in a solution close to the Kondo ion resonance frequency $\omega_s$.

The coupling between the two systems is especially important if the relaxation rate of the conduction electrons toward the Kondo ions is much faster than to the lattice and the resonance frequencies are close to one another ("bottleneck" regime):

$$\Gamma_{\sigma\sigma} \gg \Gamma_{\sigma L}, |\omega_s - \omega_\sigma|. \tag{29}$$

In the case of an isotropic system and equal Larmor frequencies ($g_\| = g_\perp = g_\sigma$) the solutions of (24) give the well known result [18, 20]: the ESR linewidth in the bottleneck regime is greatly narrowed due to conservation of the total magnetic moment (its operator commutes with the isotropic Kondo interaction and the latter disappears from the effective relaxation rate). In the opposite case of a strongly anisotropic Kondo interaction our results in the second order do not show any sufficient narrowing of the ESR linewidth in the bottleneck regime. The Kondo effect could change the situation but in this case the third order logarithmic terms in (28) become large and we need to improve the standard perturbation technique.

## 4. The renormalized relaxation rate

To develop a satisfactory theory in the low temperature regime one has to remove the logarithmic divergences by summing somehow the higher order terms in the perturbation expansion. In 1965 Abrikosov [21] carried out a summation of the leading logarithmic terms for the resistivity applying the Feynman diagrams technique to the s-d model. Later Anderson [22] proposed another method known as "Poor Man's Scaling" that allows one to extend the lowest order perturbation results, and effectively sums the leading order logarithmic terms. We use here the scaling approach because a considerable number of Kondo couplings makes it difficult to work with the diagrams technique.

The main idea of the "Poor Man's Scaling" approach is to take into account the effect of the high energy excitations on the low energy physics by a renormalization of coupling constants. We divide the conduction electrons band into the low and high energy states

$$0 < |\varepsilon_{\mathbf{k}}| < W', \quad W' < |\varepsilon_{\mathbf{k}}| < W, \tag{30}$$

where $W$, $W'$ are the initial and the running bandwidth, respectively. The projection of the original Kondo interaction $H_{s\sigma}$ (13) on to the low energy states yields a new Hamiltonian $H'_{s\sigma}$ with renormalized couplings $J'_{\alpha\beta}$. It is convenient to introduce the dimensionless parameters $U_{\alpha\beta} = (\rho J_{\alpha\beta})'$, with the symbol $(...)'$ marking all new quantities. The evolution of the renormalized parameters $U_{\alpha\beta}$ with $W'$ is described with a set of equations derived to the second order in $U_{\alpha\beta}$:

$$\frac{\partial U_{\alpha\alpha'}}{\partial t} = -\frac{1}{2} \sum_{\beta\gamma\beta'\gamma'} \varepsilon_{\alpha\beta\gamma} \varepsilon_{\alpha'\beta'\gamma'} U_{\beta\beta'} U_{\gamma\gamma'}. \tag{31}$$

Here $\varepsilon_{\alpha\beta\gamma}$ denotes an antisymmetric third-rank tensor, $t = \ln(W'/W)$; $\alpha$, $\beta$, $\gamma$ run over $x$, $y$, $z$. The initial conditions for the system (31) are the "bare" parameters of the Kondo interaction,

$$U_{\alpha\beta}(0) = \rho J_{\alpha\beta} \tag{32}$$

with $J_{\alpha\beta}$ defined by (16).

The solution of equations (31) can be written in the form

$$\begin{aligned}
U_{xx} &= U_\perp, \\
U_{yy} &= U_\perp \cos\theta \cos\psi + U_\parallel \sin\theta \sin\psi, \\
U_{zz} &= U_\perp \sin\theta \sin\psi + U_\parallel \cos\theta \cos\psi, \\
U_{yz} &= U_\perp \cos\theta \sin\psi - U_\parallel \sin\theta \cos\psi, \\
U_{zy} &= U_\perp \sin\theta \cos\psi - U_\parallel \cos\theta \sin\psi, \\
U_{xy} &= U_{yx} = U_{xz} = U_{zx} = 0,
\end{aligned} \tag{33}$$

$$U_\parallel = \bar{J} \cot\varphi, \quad U_\perp = \bar{J}/\sin\varphi, \tag{34}$$

where $\psi$ is defined by (18); $U_\parallel$, $U_\perp$ will represent the renormalized parameters of the Kondo interaction if the quantization axis is not rotated, $\bar{J} = \rho J \sqrt{g_\perp^2 - g_\parallel^2}$, $\varphi = \bar{J} \ln(W'/T_{GK})$. The abbreviation "GK" indicates the ground Kramers state and $T_{GK}$ is a characteristic temperature given as follows:

$$T_{GK} = W \exp\left[-\frac{1}{\bar{J}} \operatorname{arccot}\left(\frac{g_\parallel}{\sqrt{g_\perp^2 - g_\parallel^2}}\right)\right]. \tag{35}$$

However, the expressions just derived should be treated with care. One can see that a set of renormalized couplings (33) includes a new parameter $U_{zy}$ which is absent among the parameters of the original Hamiltonian (16), as if the model would be nonrenormalizable. On the other hand we keep in mind the anisotropy of the Zeeman energy to be the same as that of the Kondo interaction, the property being likely independent of scaling procedure. It means that $g$-factors as well as the Kondo couplings are affected by the renormalization. The relation $J_\parallel / J_\perp = g_\parallel / g_\perp$ (see (5)) still holds to result, converting, after renormalization, to $U_\parallel / U_\perp = g'_\parallel / g'_\perp = \cos\varphi$. The latter immediately leads to the renormalizability of our model: $U_{zy}(t) = 0 = U_{zy}(0)$. Note, that the angle $\psi$ is also renormalized following the equation $\cot\psi' = \cos\varphi \cot\theta$. With this correction, the expressions (33) transform to

$$U_{xx} = \bar{J} / \sin\varphi,$$
$$U_{yy} = U_{xx} \cos\varphi \Big/ \sqrt{\sin^2\theta + \cos^2\theta\cos^2\varphi},$$
$$U_{zz} = U_{xx} \sqrt{\sin^2\theta + \cos^2\theta\cos^2\varphi}, \qquad (36)$$
$$U_{yz} = U_{xx} \sin\theta\cos\theta\sin^2\varphi \Big/ \sqrt{\sin^2\theta + \cos^2\theta\cos^2\varphi},$$
$$U_{zy} = U_{xy} = U_{yx} = U_{xz} = U_{zx} = 0.$$

The characteristic temperature $T_{GK}$ is obviously independent of initial and running values of the parameters, representing a universal energy scale to govern all low temperature physics:

$$T_{GK}(\rho J_{\alpha\beta}, W) = T_{GK}((\rho J_{\alpha\beta})', W'). \qquad (37)$$

All physical quantities can be expressed in terms of the ratios $T/T_{GK}$, $B/T_{GK}$. By extending the scaling procedure down to the effective bandwidth $W' = T$ we obtain the temperature dependence of renormalized Kondo couplings:

$$\varphi(T) = \bar{J} \ln(T/T_{GK}). \qquad (38)$$

Note, that in the isotropic case ($g_\parallel = g_\perp = g$) equation (35) gives the standard result $T_{GK} = W e^{-\frac{1}{\rho J g}}$.

The temperature dependent parameters are readily seen to be logarithmically divergent at $T \to T_{GK}$ ($\varphi \ll 1$): $U_{\alpha\beta}(T) \sim 1/\ln(T/T_{GK})$. It means that the conduction electron bandwidth can only be reduced to $W' \sim T_{GK}$, the value at which the perturbative scaling approach begins to break down. Consequently, all results derived by this method are only valid for temperatures well above $T_{GK}$.

Using the expressions (36) for the temperature dependent Kondo couplings we find the renormalized kinetic coefficients

$$\Gamma_{ss}(\omega) = \pi \frac{\omega}{\omega_s} \bar{J}^2 T \left[\cot^2\varphi + \frac{1}{2} + \frac{\sin^2\theta}{4(\sin^2\theta + \cos^2\theta\cos^2\varphi)}\right],$$

$$\Gamma_{\sigma\sigma}(\omega) = \pi \frac{\omega}{\omega_s} \bar{J}^2 \frac{g_s}{2g_\sigma \rho(T+T_W)} T \left[\cot^2\varphi + \frac{1}{2} + \frac{\sin^2\theta}{4}\right], \qquad (39)$$

$$\Gamma_{\sigma s}(\omega) = \pi \frac{\omega}{\omega_s} \bar{J}^2 T \left[\sqrt{\sin^2\theta + \cos^2\theta\cos^2\varphi} + \cos\varphi\right] \Big/ 2\sin^2\varphi,$$

$$\Gamma_{s\sigma}(\omega) = \Gamma_{\sigma s}(\omega) \frac{g_s}{2g_\sigma \rho(T+T_W)}.$$

One can see that all kinetic coefficients logarithmically diverge at $T \to T_{GK}$ ($\varphi \ll 1$): to the leading order in logarithmic terms they are of the form

$$\Gamma_{ss}(\omega) = \Gamma_{\sigma s}(\omega) = \pi \frac{\omega}{\omega_s} \frac{T}{\ln^2(T/T_{GK})},$$

$$\Gamma_{\sigma\sigma}(\omega) = \Gamma_{s\sigma}(\omega) = \Gamma_{ss}(\omega) \frac{g_s}{2g_\sigma \rho(T+T_W)}.$$

(40)

At first glance, these results confirm the common belief that the ESR linewidth of Kondo ions (as well as conduction electrons) is expected to be too large for its detection. However, the coupling between two systems makes the situation quite different. We solve the coupled equations (24) under the condition of the bottleneck regime (29), taking into account the renormalization of kinetic coefficients. The relaxation rate of the collective spin mode with a frequency close to the Kondo-ion resonance now follows

$$\Gamma_{coll} = \Gamma_{sL} + \Gamma_{ss}^{eff} + \Gamma_{\sigma L}^{eff};$$

$$\Gamma_{ss}^{eff} = \Gamma_{ss} - \frac{\Gamma_{s\sigma}\Gamma_{\sigma s}}{\Gamma_{\sigma\sigma}}, \qquad \Gamma_{\sigma L}^{eff} = \Gamma_{\sigma L} \frac{\Gamma_{s\sigma}\Gamma_{\sigma s}}{\Gamma_{\sigma\sigma}^2}.$$

(41)

It is interesting to analyze the asymptotic behavior of an effective Korringa relaxation rate $\Gamma_{ss}^{eff}$ and an effective relaxation rate of conduction electrons to the lattice $\Gamma_{\sigma L}^{eff}$ at $T \to T_{GK}$. In this case their expressions are essentially simplified to be written explicitly as functions of temperature and angle between the static magnetic field $B$ and the crystal symmetry axis $c$:

$$\Gamma_{ss}^{eff} = \frac{\pi}{8} \bar{J}^4 (2 - \sin^4\theta) T \ln^2(T/T_{GK}),$$

(42)

$$\Gamma_{\sigma L}^{eff} = \rho(T+T_W) \frac{2g_\sigma}{g_s} \Gamma_{\sigma L}.$$

(43)

The result is rather unusual: instead of being divergent, the effective Korringa relaxation rate is greatly reduced and goes to zero at $T \to T_{GK}$. Although the Kondo interaction is strongly anisotropic, the Kondo effect leads to the common energy scale $T_{GK}$ regulating the temperature dependence of different physical quantities. Indeed, all kinetic coefficients in (40) display a similar logarithmic temperature dependence leading to their complete mutual cancelation (to the leading order terms only!) in the collective spin mode. The effective relaxation rate of the conduction electrons to the lattice $\Gamma_{\sigma L}^{eff}$ is also greatly reduced, becoming proportional to temperature.

The angular dependence of the effective relaxation rate qualitatively agrees with the ESR linewidth as experimentally observed. $\Gamma_{ss}^{eff}$ varies monotonically for $0 < \theta < \pi/2$. The lowest and highest values of $\Gamma_{ss}^{eff}$ correspond to the perpendicular and parallel orientation, respectively. The maximum of $\Gamma_{ss}^{eff}$ is two times larger than minimum. This dependence is likely due to the partial breaking of the bottleneck regime condition (29) in the case of the parallel orientation, when the inequality $\omega_\sigma \gg \omega_s$ is fulfilled.

However, one must remember that all divergences and cancelations in relaxation rates should not be taken literally because the perturbational scaling approach is only valid for the temperatures well above $T_{GK}$. As a matter of fact, the singularities just indicate an increase of physical quantities with temperatures lowering to $T_{GK}$. Moreover, the considerations based on a single ion Kondo problem suggests that the Kondo anomalies are quenched, provided the static magnetic field is large compared with $T_{GK}$. At the temperature region $T_{GK} < T < B$ the thermodynamic functions become temperature independent and the ratio $T/T_{GK}$ can be replaced with $B/T_{GK}$. In the previous work [4] the fitting to the experimental data revealed $T_{GK} = 0.36$ K, what is by two orders of magnitude smaller than the Kondo temperature $T_K$ derived thermodynamically [23] and by transport measurements [24]. The resonance field of the ESR experiments [1, 5] is of the order of $T_{GK}$: $B \equiv 0.46$ K at X-band and $B \equiv 1.62$ K at Q-band. For these reasons, the scaling approach for the ESR measurements seems to be valid for the temperature region $T_{GK} < B < T < T_K$.

For detailed analysis of the ESR linewidth angular dependence, as well as for a comparison with experimental data also contributions other than Kondo interaction need to be taken into account. Now we merely discuss the broadening of ESR linewidth which is represented by the kinetic coefficient $\Gamma_{sL}$. An obvious contribution comes from the usual magnetic dipole-dipole interactions of the Kondo ions. The corresponding local magnetic field can be easily evaluated and gives approximately $\Delta B_{loc} \sim 1000$ Oe [25]. This value is much larger than the observed ESR linewidth in YbRh$_2$Si$_2$ for X- and Q-bands ($\Delta B_{loc} \sim 200$ Oe at $T = 5$ K). The contribution from the RKKY interaction, which becomes highly anisotropic after projection onto the ground Kramers doublet, is expected to be much larger. So, it is evident that some narrowing mechanism for these type of contributions should exist. It is well established [18] that in the bottleneck regime the broadening of the ESR line by the distribution of local fields is sufficiently reduced due to fast reorientation of the Kondo ion moment caused by the Korringa relaxation. The corresponding contribution to the linewidth can be estimated as $\langle(\Delta\nu)^2\rangle / \Gamma_{ss}$ (but not $\Gamma_{ss}^{eff}$ !), where $\langle(\Delta\nu)^2\rangle$ is the mean square distribution of the resonance frequencies due to the local fields. An additional broadening of the ESR line comes from the spin-phonon interaction of Kondo ions. It is enough to take into account the two-phonon Raman and Orbach processes at temperatures above a few K. This contribution can not be the subject for the narrowing described above.

## 5. Conclusion
We investigated the spin relaxation in the heavy fermion Kondo lattices YbRh$_2$Si$_2$ and YbIr$_2$Si$_2$ at temperatures well below thermodynamically measured Kondo temperature $T_K = 25$ K. The idea of a collective spin motion of Kondo ions with conduction electrons in presence of the Kondo effect successfully explains the ESR signal existence in these compounds. We conclude the article with a few comments.

The relaxation rate of the collective spin mode is found to be strongly dependent on the orientation of the static magnetic field to the crystal symmetry axis. Its angular dependence reflects the anisotropy of the Kondo interaction to confirm the idea of collective spin motion in the bottleneck regime. The broadening of the ESR linewidth with approaching to the parallel orientation can be attributed to the partial opening of the bottleneck due to significant difference between resonance frequencies of conduction electrons and Yb-ions. The treatment of the Kondo interaction alone provides for a qualitative agreement with experimental data on the angular dependence of the ESR linewidth. To obtain a quantitative agreement other contributions than the Kondo interaction needs be taken into account.

Another point to treat is the ratio of $g$-factors $g'_\parallel / g'_\perp = \cos\varphi$ used in derivation of the renormalized relaxation rate. One can see that at $T \to T_{GK}$ ($\varphi \ll 1$) the $g$-factors tend to be equal. This seems to contradict the experimentally observed strong anisotropy of the $g$-factor. However, the $g$-factors mentioned above are not the same as the experimentally observed ones. The resonance frequency of the collective spin mode is determined by the total dynamical susceptibility as a real part of its pole, whereas the $g$-factors in the Zeeman energy are just the matrix elements of the total electronic moment of an Yb-ion (see (4)). The equality $\left(g_\parallel^{Yb}\right)' = \left(g_\perp^{Yb}\right)'$ does not mean a similar relation in the case of resonance frequency of the collective mode, what is the quantity to be compared with experimental data.

Finally, we discuss the new characteristic temperature $T_{GK} = 0.36$ K revealed by the perturbational scaling approach and describing the *g*-data [4]. This parameter represents a common energy scale to regulate the behavior of the different kinetic coefficients. The scaling approach is valid at temperatures above $T_{GK}$. Hence, our theory is well suitable at the temperature region of the ESR measurements. At present the relation between $T_{GK}$ and $T_K$ remains an open question. We can only suggest them to be different energy scales, the former controlling the magnetic relaxation, and the latter is probably responsible for transport and thermal properties.

**Acknowledgments**
This work was supported by the Volkswagen Foundation (I/84689) and partially by the Ministry of Education and Science of the Russian Federation. The authors are grateful to J. Sichelschmidt for the fruitful discussions and for help in the preparation of the manuscript.

**Appendix**
Let us introduce a functional of two external random fields

$$V = \sum_1 \left(-\mathbf{v}_1 \mathbf{S}_1 + \mathbf{w}_1 \boldsymbol{\sigma}_1\right). \tag{A.1}$$

Here $1 \equiv (i_1, \tau_1)$, $i_1$ labels a lattice site, $\tau_1$ is imaginary time at the interval $0 < \tau_1 < \beta$, $\beta = 1/k_B T$; $\mathbf{v}_1$, $\mathbf{w}_1$ are external random fields, $\sum_1 \equiv \sum_{i_1} \int_0^\beta$; $\mathbf{S}_1$, $\boldsymbol{\sigma}_1$ represent spin operators of Yb-ion and conduction electrons, respectively:

$$\mathbf{S}_1 = e^{\tau_1 H} \mathbf{S}_{i_1} e^{-\tau_1 H}, \quad \boldsymbol{\sigma}_1 = e^{\tau_1 H} \boldsymbol{\sigma}_{i_1} e^{-\tau_1 H}. \tag{A.2}$$

The spin operators $\mathbf{S}_1$, $\boldsymbol{\sigma}_1$ can be written in terms of variational derivatives of the functional (A.1) with respect to the external random fields:

$$\frac{\delta V}{\delta v_1^\alpha} = -S_1^\alpha, \quad \frac{\delta V}{\delta w_1^\alpha} = \sigma_1^\alpha, \quad \alpha = x, y, z. \tag{A.3}$$

For any set of time dependent operators we define the temperature Green function

$$\langle T A(\tau_1) B(\tau_2)...\rangle = \mathrm{Sp}\left(e^{-\beta H} T A(\tau_1) B(\tau_2)...e^{-V}\right) \Big/ \mathrm{Sp}\left(e^{-\beta H} T e^{-V}\right), \tag{A.4}$$

where *T* is the time ordering operator.

Functional derivation allows one to find composite averages, for example,

$$\frac{\delta \langle TA \rangle}{\delta v_1^\alpha} = \langle TAS_1^\alpha \rangle - \langle TA \rangle \langle TS_1^\alpha \rangle,$$

$$\frac{\delta \langle TA \rangle}{\delta w_1^\alpha} = -\langle TA\sigma_1^\alpha \rangle + \langle TA \rangle \langle T\sigma_1^\alpha \rangle. \tag{A.5}$$

The Green functions (A.4) follow equation of motion

$$\frac{\partial \langle TA(\tau) \rangle}{\partial \tau} = \langle T\dot{A}(\tau) \rangle - \langle T\{A(\tau), V\}_+ \rangle, \tag{A.6}$$

$$\frac{\partial \langle TA(\tau_1) B(\tau_2) \rangle}{\partial \tau_1} = \langle T\{A(\tau_1), B(\tau_2)\}_\eta \rangle + \langle T\dot{A}(\tau_1) B(\tau_2) \rangle - \langle T\{A(\tau_1), V\}_+ B(\tau_2) \rangle \tag{A.7}$$

with $\{A(\tau_1), B(\tau_2)\}_\eta = \delta(\tau_1 - \tau_2)\left(A(\tau_1) B(\tau_2) - \eta B(\tau_2) A(\tau_1)\right)$, $\eta = +1, -1$ in the cases of Bose- and Fermi operators, respectively.

To study the ESR response we need four two-spin Green functions

$$G_{12}^{\sigma_1\sigma_2} = -\langle T c_{1\sigma_1} c_{2\sigma_2}^+ \rangle,$$
$$D_{12}^{\alpha\beta} = -\langle T S_1^\alpha S_2^\beta \rangle_c,$$
$$A_{12}^{\alpha\beta} = -\langle T \sigma_1^\alpha \sigma_2^\beta \rangle_c,$$
$$B_{12}^{\alpha\beta} = -\langle T \sigma_1^\alpha S_2^\beta \rangle_c.$$
(A.8)

Here $\langle TAB \rangle_c = \langle TAB \rangle - \langle TA \rangle \langle TB \rangle$, $\sigma_1$, $\sigma_2$ denote the orientation of conduction electron spin. The Green functions (A.8) are coupled with obvious relations reducing the problem to calculation of two independent functions $G$, $D$:

$$A_{12}^{\alpha\beta} = \frac{\delta \langle T\sigma_1^\alpha \rangle}{\delta w_2^\beta} = \sum_{\lambda\mu} s_{\lambda\mu}^\alpha \frac{\delta G_{11}^{\mu\lambda}}{\delta w_2^\beta}, \qquad B_{12}^{\alpha\beta} = -\frac{\delta \langle T\sigma_1^\alpha \rangle}{\delta v_2^\beta} = -\sum_{\lambda\mu} s_{\lambda\mu}^\alpha \frac{\delta G_{11}^{\mu\lambda}}{\delta v_2^\beta}.$$
(A.9)

At the end of all calculations the external random fields $\mathbf{v}$, $\mathbf{w}$ are set to zero, and Green functions can be written in terms of their Fourier transforms

$$f_{12} = \frac{1}{N\beta} \sum_{\mathbf{k},n} e^{i\mathbf{k}(\mathbf{r}_1-\mathbf{r}_2)} e^{-i\omega_n(\tau_1-\tau_2)} f(\mathbf{k}, i\omega_n),$$
(A.10)

where $\omega_n = 2\pi n/\beta$, $\omega_n = \pi(2n+1)/\beta$ in the cases of Bose- and Fermi operators, respectively.

Using (A.7), (A.8) we obtain an equation of motion for the one-electron Green function $G_{12}^{\sigma_1\sigma_2}$

$$F_{11'}^{\sigma_1\sigma_1'} G_{1'2}^{\sigma_1'\sigma_2} = -\delta_{\sigma_1\sigma_2}\delta_{12} - \delta_{11'} J_{\alpha\beta} s_{\sigma_1\sigma_1'}^\alpha \left( \langle TS_1^\beta \rangle + \frac{\delta}{\delta v_1^\beta} \right) G_{1'2}^{\sigma_1'\sigma_2},$$
(A.11)

where

$$F_{11'}^{\sigma_1\sigma_1'} = \delta_{\sigma_1\sigma_1'}\delta_{11'}\left(\frac{\partial}{\partial \tau_1} + \varepsilon^{\sigma_1}\right) + \delta_{\sigma_1\sigma_1'} t_{11'} + \delta_{11'}(\mathbf{w}_1 \mathbf{s}_{\sigma_1\sigma_1'}),$$
(A.12)

$\delta_{11'} = \delta_{i_1i_1'}\delta(\tau_1-\tau_{1'})$, $t_{11'} = t_{i_1i_1'}\delta(\tau_1-\tau_{1'})$; $s_{\sigma_1\sigma_1'}^\alpha$, $\varepsilon^\lambda$, $J_{\alpha\beta}$ are defined in (9), (15), (16), a sum over identical indices is implied.

The equation of motion for the localized spin Green function is derived in a similar way:

$$\left(\lambda \frac{\partial}{\partial \tau_1} - g_s B + v_1^z\right) D_{12}^{\lambda\bar\lambda} = -\delta_{12} \langle TS_1^z \rangle + v_1^\lambda D_{12}^{z\bar\lambda} +$$
$$+ J_{\alpha z}\left[ D_{12}^{\lambda\bar\lambda} - \langle TS_1^\lambda \rangle \frac{\delta}{\delta v_2^\lambda} + \frac{\delta^2}{\delta v_1^{\bar\lambda} \delta v_2^\lambda} \right] s_{\lambda\mu}^\alpha G_{11}^{\mu\lambda} - J_{\alpha\lambda}\left[ D_{12}^{z\bar\lambda} - \langle TS_1^z \rangle \frac{\delta}{\delta v_2^\lambda} + \frac{\delta^2}{\delta v_1^z \delta v_2^\lambda} \right] s_{\lambda\mu}^\alpha G_{11}^{\mu\lambda} + \quad (A.13)$$
$$+ I_{11'}^{z\alpha}\left[ \langle TS_1^\lambda \rangle D_{1'2}^{\alpha\bar\lambda} + \langle TS_{1'}^\alpha \rangle D_{12}^{\lambda\bar\lambda} - \frac{\delta}{\delta v_{1'}^\alpha} D_{12}^{\lambda\bar\lambda} \right] - I_{11'}^{z\alpha}\left[ \langle TS_1^z \rangle D_{1'2}^{\alpha\bar\lambda} + \langle TS_{1'}^\alpha \rangle D_{12}^{z\bar\lambda} - \frac{\delta}{\delta v_{1'}^\alpha} D_{12}^{z\bar\lambda} \right].$$

Here $I_{11'} = I_{i_1i_1'}\delta(\tau_1-\tau_{1'})$, $J_{\alpha\lambda} = (J_{\alpha x} + i\lambda J_{\alpha y})/\sqrt{2}$, $I^{\lambda\alpha} = (I^{x\alpha} + i\lambda I^{y\alpha})/\sqrt{2}$, $\lambda = +1, -1$, $\bar\lambda = -\lambda$.

Equations of motion are solved by a perturbational expansion on Kondo couplings $J_{\alpha\beta}$. The treatment of both equations (A.11) and (A.13) are basically similar to each other, but the latter is much more complicated, so we shall not write it out explicitly.

It is convenient to introduce a self energy for the one electron Green function

$$G^{-1} = G_0^{-1} - \Sigma,$$
(A.14)

$G_0$ is the Green function in the absence of the Kondo interaction, indices 1, 2, $\sigma_1$, $\sigma_2$ are dropped. Combining (A.12) with (A.14) we immediately obtain

$$G_0^{-1} = -F, \tag{A.15}$$

$$\Sigma_{12}^{\sigma_1\sigma_2} = \delta_{12} J_{\alpha\beta} s_{\sigma_1\sigma_2}^{\alpha} \langle TS_1^{\beta} \rangle + J_{\alpha\beta} s_{\sigma_1\lambda}^{\alpha} G_{11'}^{\lambda\lambda'} \frac{\delta \Sigma_{1'2}^{\lambda\sigma_2}}{\delta v_1^{\beta}}. \tag{A.16}$$

An iterative process with equation (A.16) yields $\Sigma = \Sigma_1 + \Sigma_2 + \Sigma_3 + O(J^4)$, where

$$\begin{aligned}
(\Sigma_1)_{12}^{\sigma_1\sigma_2} &= \delta_{12} J_{\alpha\alpha'} s_{\sigma_1\sigma_2}^{\alpha} \langle TS_1^{\alpha'} \rangle, \\
(\Sigma_2)_{12}^{\sigma_1\sigma_2} &= -J_{\alpha\alpha'} J_{\beta\beta'} s_{\sigma_1\lambda}^{\alpha} s_{\lambda'\sigma_2}^{\beta} G_{12}^{\lambda\lambda'} D_{21}^{\beta'\alpha'}, \\
(\Sigma_3)_{12}^{\sigma_1\sigma_2} &= -J_{\alpha\alpha'} J_{\beta\beta'} J_{\gamma\gamma'} s_{\sigma_1\lambda}^{\alpha} s_{\lambda'\mu'}^{\beta} s_{\mu\sigma_2}^{\gamma} G_{11'}^{\lambda\lambda'} G_{1'2}^{\mu\mu} \frac{\delta D_{1'1}^{\beta'\alpha'}}{\delta v_{2'}^{\gamma'}},
\end{aligned} \tag{A.17}$$

the variational derivative $\dfrac{\delta D}{\delta v}$ is a three spin Green function:

$$-\frac{\delta D_{12}^{\alpha\beta}}{\delta v_3^{\gamma}} = \langle TS_1^{\alpha} S_2^{\beta} S_3^{\gamma} \rangle_c = \langle TS_1^{\alpha} S_2^{\beta} S_3^{\gamma} \rangle + \langle TS_1^{\alpha} \rangle D_{23}^{\beta\gamma} + \langle TS_2^{\beta} \rangle D_{13}^{\alpha\gamma} + \langle TS_3^{\gamma} \rangle D_{12}^{\alpha\beta} + \langle TS_1^{\alpha} \rangle \langle TS_2^{\beta} \rangle \langle TS_3^{\gamma} \rangle.$$

Putting $V = 0$ we obtain the first order self energy

$$(\Sigma_1)_{12}^{\lambda\lambda} = \delta_{12} \lambda J_{zz} \langle S^z \rangle / 2, \qquad (\Sigma_1)_{12}^{\lambda\bar{\lambda}} = \delta_{12} J_{\bar{\lambda}z} \langle S^z \rangle / \sqrt{2}, \qquad \langle S^{\alpha} \rangle \equiv \langle TS_1^{\alpha} \rangle_{V=0}. \tag{A.18}$$

Diagonal part of $\Sigma_1$ gives the shift of the conduction electrons resonance frequency, which, after Fourier transform, can be included into the zero order Green function

$$G_0^{\lambda\lambda}(\mathbf{k}, i\omega_n) = \frac{1}{i\omega_n - \varepsilon_{\mathbf{k}}^{\lambda}}, \tag{A.19}$$

where $\omega_n = \pi(2n+1)/\beta$, $\varepsilon_{\mathbf{k}}^{\lambda} = t(\mathbf{k}) - \mu + \lambda\omega_{\sigma}/2$, $\omega_{\sigma} = g_{\sigma}B + J_{zz}\langle S^z \rangle$. Exact expressions for $G^{\sigma_1\sigma_2}(\mathbf{k}, i\omega_n)$ are of the form

$$G^{\lambda\lambda} = \left[ (G_0^{-1})^{\lambda\lambda} - \Sigma^{\lambda\lambda} - \frac{\Sigma^{\lambda\bar{\lambda}} \Sigma^{\bar{\lambda}\lambda}}{(G_0^{-1})^{\bar{\lambda}\bar{\lambda}} - \Sigma^{\bar{\lambda}\bar{\lambda}}} \right]^{-1}, \qquad G^{\lambda\bar{\lambda}} = \frac{\Sigma^{\lambda\bar{\lambda}}}{(G_0^{-1})^{\bar{\lambda}\bar{\lambda}} - \Sigma^{\bar{\lambda}\bar{\lambda}}} G^{\lambda\lambda}. \tag{A.20}$$

Substitution of (A.17) into (A.20) gives the non-diagonal Green function as follows

$$G_{12}^{\lambda\bar{\lambda}} = \frac{J_{\bar{\lambda}z}}{\sqrt{2}} \frac{\langle S^z \rangle}{\omega_{\sigma}} \sum_{\mu} \mu G_{12}^{\mu\mu} + O(J^2). \tag{A.21}$$

A similar treatment of equation (A.13) yields the localized spin Green function to the first order in $J$, the RKKY interaction taken into account in molecular field approximation only. The result is

$$D^{\lambda\bar{\lambda}}(\mathbf{k}, i\omega_n) = \frac{\lambda \langle S^z \rangle}{i\omega_n - \lambda\omega_s} + O(J^2), \qquad D^{zz}(\mathbf{k}, i\omega_n) = -\frac{\beta}{4} \delta_{n0} + O(J^2), \tag{A.22}$$

where $\omega_n = 2\pi n/\beta$, $\omega_s$ is given by (26).

Two-spin Green functions $A$, $B$ are derived from (A.9) via the obvious relations

$$\begin{aligned}
\delta G &= -G \delta G_0^{-1} G + G \delta \Sigma G, \\
\frac{\delta (G_0^{-1})_{12}^{\lambda\mu}}{\delta w_3^{\alpha}} &= -\delta_{12} \delta_{23} s_{\lambda\mu}^{\alpha}, \qquad \frac{\delta G_0^{-1}}{\delta v} = 0.
\end{aligned} \tag{A.23}$$

The calculation of all Green functions up to the third order in $J$ results in the system of coupled equations (24). Here we set out explicit expressions for $A$, $B$ to the first order in $J$ only, since their higher order terms are too lengthy.

$$A_{12}^{zz} = \frac{1}{4}\sum_\lambda G_{12}^{\lambda\lambda} G_{21}^{\lambda\lambda} + O(J^2), \qquad A_{12}^{\lambda\bar\lambda} = \frac{1}{2} G_{12}^{\bar\lambda\bar\lambda} G_{21}^{\lambda\lambda} + O(J^2),$$

$$A_{12}^{\lambda z} = A_{21}^{z\lambda} = J_{\lambda z}\frac{\langle S^z\rangle}{\omega_\sigma}\left(A_{12}^{zz} - A_{12}^{\lambda\bar\lambda}\right) + O(J^2), \qquad A_{12}^{\lambda\lambda} = O(J^2). \qquad (A.24)$$

$$B_{12}^{\alpha\beta} = J_{\alpha'\beta'} A_{11'}^{\alpha\alpha'} D_{1'2}^{\beta'\beta} + O(J^2). \qquad (A.25)$$